\documentclass[aps,prl,showpacs,superscriptaddress,amssymb,twocolumn]{revtex4-2}
\usepackage{blindtext}
\usepackage{appendix} 
\usepackage[colorlinks=true, allcolors=blue]{hyperref}
\usepackage{bm,amsmath}
\usepackage{dcolumn}
\usepackage{braket}
\usepackage{mathtools}
\usepackage{physics}
\usepackage{algorithm2e}
\usepackage[utf8]{inputenc}
\usepackage[T1]{fontenc}
\usepackage{graphicx}

\begin{document}

\rightline{\tt }

\vspace{0.2in}

\title{Quantum algorithm for Wang-Landau sampling}

\author{Garrett T. Floyd, David P. Landau, Michael R. Geller}
\affiliation{Department of Physics and Astronomy, University of Georgia, Athens, Georgia 30602, USA and
Center for Simulational Physics, University of Georgia, Athens, Georgia 30602, USA}

\date{\today}

\begin{abstract}

It has been shown that the Metropolis algorithm can be implemented on quantum computers in a way that avoids the sign problem. However, flat histogram techniques are often preferred as they don’t suffer from the same limitations that afflict Metropolis for problems of real-world interest and provide a host of other benefits. In particular, the Wang-Landau method is known for its efficiency and accuracy. In this work we design, implement, and validate a quantum algorithm for Wang-Landau sampling, greatly expanding the range of quantum many body problems solvable by Monte Carlo simulation.
\end{abstract}

\maketitle

\section{Introduction}
\label{introduction section}
Monte Carlo techniques are now ubiquitous in physics and related disciplines because of their ability to calculate finite-temperature equilibrium properties  \cite{david_p_landau_guide_2015}. The simplest and most well-known of these techniques is the ``traditional'', Markov chain Metropolis algorithm. Nonetheless, traditional Monte Carlo algorithms have found limited applicability to quantum systems due to the infamous sign problem. On the other hand, quantum computers promise to efficiently simulate a broad range of quantum models and provide vastly increased speed and accuracy compared to classical algorithms.

It has been shown by Temme {\it et al.}~\cite{temme_quantum_2011} that the Metropolis algorithm can be adapted for quantum computers while avoiding the sign problem. Since then interest in the quantum simulation of thermal states has intensified. This has given rise to a number of advances in quantum computer based Markov chain methods  \cite{nelson_high-quality_2022,somma_quantum_2008,lemieux_efficient_2020, moussa_measurement-based_2022, terhal_problem_2000, wocjan_quantum_2009,chiang_quantum_2010,wild_quantum_2021,endo_quantum_2020,montanaro_quantum_2015, yung_quantumquantum_2012, arunachalam_simpler_2022} as well as new techniques such as those based on variational methods \cite{motta_determining_2020, chen_hybrid_2020, matsumoto_calculation_2022,verdon_quantum_2019,sagastizabal_variational_2021,wu_variational_2019,chowdhury_variational_2020,shingu_boltzmann_2021,wang_variational_2021}, and others  \cite{lu_algorithms_2021,bassman_computing_2021,brandao_finite_2019,cohn_minimal_2020,martyn_product_2019,wild_quantum_2021-1,ge_rapid_2016,poulin_sampling_2009,wu_estimating_2021,francis_many_2020,bilgin_preparing_2010,chowdhury_quantum_2016,van_apeldoorn_quantum_2020,alhambra_locally_2021}. As an example, while no rigorous complexity bounds are known for the classical or quantum Metropolis algorithms, many of the works based on Markov chain methods have shown that an additional quadratic speedup of hitting times can be achieved. This is often associated with an optimal choice of quantum walk operator \cite{yung_quantumquantum_2012,lemieux_efficient_2020,wocjan_quantum_2009,wild_quantum_2021,wild_quantum_2021-1,montanaro_quantum_2015, arunachalam_simpler_2022,chowdhury_quantum_2016,van_apeldoorn_quantum_2020,an_quantum-accelerated_2021,holbrook_quantum_2022,herbert_quantum_2021}. 

Yet despite the popularity of the Metropolis and other simple Markov chain based algorithms, these approaches suffer from considerable defects. It is well known that Metropolis algorithms exhibit critical slowing down at second order phase transitions and can become trapped at low temperatures when sampling complex free energy landscapes. Metropolis algorithms also require separate runs for each temperature investigated. Furthermore, not all thermodynamic quantities of interest, e.g. entropy and free energy, can be calculated directly from the raw data provided from Metropolis.  These limitations often necessitate the use of advanced post-Metropolis simulation methods. 

 One such technique is the Wang-Landau Monte Carlo algorithm \cite{wang_efficient_2001,wang_determining_2001}. The Wang-Landau algorithm is known to sample complex free energy landscapes while maintaining applicability and simplicity. The Wang-Landau algorithm works by performing a random walk in energy space in a manner which depends on the current estimate of the density of states (DOS), which is updated at each step of the walk. As the walk transition probability discourages visiting energies with higher DOS estimates, it is better at exploring the entire phase space than the Metropolis method. Due to the fact that the algorithm samples a temperature independent DOS, properties of the system can be estimated simultaneously at all temperatures, avoiding trapping by free energy barriers.

Implementing a Wang-Landau algorithm on a quantum computer is challenging, however. Unlike the Metropolis algorithm whose random walk depends solely on the energies of the current and proposed wave functions, the Wang-Landau algorithm is dependent on the current estimate of the DOS entries for the current and proposed energies. It therefore requires storing, accessing, and updating the DOS, $g(E)$. While the excellent performance of the Wang-Landau algorithm has been convincingly demonstrated, at least on classical models, rigorous bounds on the convergence and accuracy are not available.

To demonstrate the feasibility and advantages of Wang-Landau sampling with a quantum computer, a proof of concept is critical. Given the considerably increased number of qubits needed to incorporate the DOS into the Monte Carlo algorithm, the extreme circuit depth required of Monte Carlo algorithms, and the exponential scaling cost of simulating additional qubits on a classical computer, providing a viable proof of concept requires some simplification. Here we design and simulate a quantum Wang-Landau algorithm that overcomes the sign problem and allows increased performance  compared to Metropolis.

 
To begin any Wang-Landau algorithm, a Hamiltonian and state space is specified. Here we denote the classical state of the system by a vector ${\bf X}$ in ${\mathbb R}^n$ or a subset of ${\mathbb R}^n$, for a (typically large) integer $n$ proportional to the number of degrees of freedom. For a system of $N$ Ising spins, ${\bf X} \in \{ -1, +1 \}^N$. Next, a fixed set of move operators $\{ C_m \}_{m=1}^M$ is chosen, with $M$ independent of $n$. The move operators are randomly chosen and applied to ${\bf X}$. The energy range over which the system is simulated is then partitioned into bins which are labeled by an index $i$. An initial guess for the (logarithm of the) relative DOS $\ln(g(E))$ is then made, and a histogram is initialized with $h(E_i) = 0  \  \forall$ i. Next a modification factor $f > 1$ and reduction factor $0 < \gamma < 1$ are chosen. A criterion determining the acceptable flatness of the histogram (flatness criterion) needed to move between iterations and the total number of iterations (max number of iterations) to be performed in the simulation is also defined.

For reference, the classical Wang-Landau algorithm is briefly summarized in Algorithm \ref{alg:Classical Wang-Landau Algorithm}. The output of any Wang-Landau algorithm is a $g(E)$ which approximates the (absolute or relative) DOS of the model. Thermodynamic quantities such as the internal energy and entropy are then obtained directly from the $g(E)$.

\RestyleAlgo{ruled}
\begin{algorithm}[hbt!]
\caption{Classical Wang-Landau Algorithm}\label{alg:Classical Wang-Landau Algorithm}
\KwData{Energy function,\\
Set of move operators to be randomly selected $\{ C_m \}_{m=1}^M$,\\
Number of energy bins $\ell$,\\
Initial guess for the logarithm of the relative DOS $\ln(g(E))$,\\
Initial histogram $h(E_i) = 0  \  \forall$ i,\\
Modification factor $f > 1$,\\
Modification reduction factor $0 < \gamma < 1$,\\
Histogram flatness criterion (flatness criterion),\\
Total number of rounds (max number of rounds).}
Set round number = 0\\
\While{\normalfont{round number < max number of rounds}}{
\For{ \normalfont{a predetermined number of steps}}
{
apply a randomly chosen move $C_m $\\
${\bf X}_i \xrightarrow{C_m} {\bf X}_j$\\
draw a random number $\in[0,1]$\\
\uIf{\normalfont{random number} < min(1,$e^{\ln(g(E_i))-\ln(g(E_j))}$)}{
Set $\ln(g(E_j)) = \ln(g(E_j)) + \ln(f)$\\
Set $h(E_j) = h(E_j) + 1$\\
Set ${\bf X}_i = {\bf X}_j$\\
}
\Else
{
Set $\ln(g(E_i)) = \ln(g(E_i)) + \ln(f)$\\
Set $h(E_j) = h(E_j) + 1$\\
Set ${\bf X}_i = {\bf X}_i$\\
}}
Perform histogram check (Algorithm \ref{alg:Check_Histogram_Subalgorithm})\\
}
\end{algorithm}

\begin{algorithm}[hbt!]
\caption{Check Histogram Subalgorithm}\label{alg:Check_Histogram_Subalgorithm}
Calculate average histogram entry ${\bar h}$\\
${\bar h} = \sum_{i=1}^\ell\frac{h(E_i)}{\ell}$\\
\uIf{$\exists$ i : $h(E_i)<$ \normalfont{(flatness criterion)}${\bar h}$ or $h(E_i)>$ \normalfont{(2-flatness criterion)}${\bar h}$}
{pass\\}
\Else
{
$\forall$ i Set $h(E_i) = 0$\\
Set $f = f^\gamma$\\
Set round number = round number + 1\\
}
\end{algorithm}

With the classical algorithm as a starting point, we then take inspiration from the method employed by Temme \cite{temme_quantum_2011}. As a review, we note that in Ref.~\cite{temme_quantum_2011} the authors perform Metropolis Monte Carlo in the following way: They begin by allocating a quantum circuit with 4 registers. The first register is used to contain the wavefunction of the system itself, the second and third are used to contain the pre- and post-update energy estimations, respectively, and the last register is a single qubit used to encode the acceptance/rejection probability.

Starting from an initial eigenstate and energy, a random unitary move update is chosen from a universal gate set $\{ C_m \}_{m=1}^M$ that is constructed to fulfill detailed balance. In general this will lead to the state register being in a superposition of eigenstates. Phase estimation is then used to get an estimate of the energy, which is used to compute and encode the acceptance probability in the last qubit for each branch of the superposition. The last qubit is then measured to see if the move is accepted or not.

If accepted, the new energy is measured from register 3, collapsing the wavefunction to a new eigenstate. The registers are then prepared for the next step. If the move is rejected, a procedure is undertaken to return the system wavefunction to an eigenstate which has the same energy as the pre-update eigenstate. Following this prescription, physical quantities can be estimated from the list of energies generated. However, from an efficiency standpoint the rejection procedure is often lengthy and computationally expensive compared to the rest of the algorithm. It has also faced criticism for the way it uses inverse phase estimation \cite{rall_faster_2021}.

To understand this, suppose that a step of the rejection procedure should use inverse phase estimation to return the state of the third register to the all zero state, for example. Note that the qubit used to store the acceptance-rejection probabilities was entangled with the third register directly before its measurement. This causes the probability amplitudes of each branch of the superposition to be altered by the measurement outcome of the final qubit. This, in turn, causes the probability of the inverse phase estimation returning the third register to the all zero state to be less than one in the case of finite-precision phase estimation. If this occurs it causes all further phase estimations in the algorithm to fail, corrupting the algorithm. This may possibly be alleviated in the limit of a large number of phase estimation qubits, but it has also recently been argued that coherent uncomputation of phase estimation after measurement cannot be generally guaranteed to succeed \cite{rall_faster_2021}. Here we employ an alternative that both remedies this issue and greatly improves the speed.

Consider the state prepared in step 2 of Yung and Aspuru-Guzik \cite{yung_quantumquantum_2012}, which is the infinite temperature thermal state of the Hamiltonian. Namely, we use the result that
\begin{equation}
 \frac{1}{\sqrt{N}}\sum_{x=0}^{N-1}\ket{x}\ket{x}=\frac{1}{\sqrt{N}}\sum_{i=0}^{N-1}\ket{\psi_i}\ket{\Bar{\psi_i}}, 
 \end{equation}
where
\begin{equation}
\ket{\Bar{\psi_i}} =\sum_{x=0}^{N-1}\braket{\psi_i}{x}\ket{x},
 \end{equation}
with $\ket{x}$ a computational basis states and the $\ket{\psi_i}$ are eigenstates of the simulated Hamiltonian. If phase estimation is performed on the first register and measured, the effect is that there is an equal probability of selecting any eigenstate of the Hamiltonian. Using preparation of this state as a wavefunction update subroutine, we note that for any initial eigenstate, a move to any other eigenstate is proposed with the same probability. There is then no need to re-prepare the initial system wavefunction in the case of rejection as the energy and all compatible observables have already been recorded and the choice of move update has no dependence on the initial wave function. Since this global move is able to select any eigenstate and does so with equal probability, it also satisfies detailed balance requirements. The need for uncomputation and wave function re-initialization can then be removed at the cost of doubling the number of qubits necessary for the system wavefunction.

It should also be noted that this approach allows the acceptance-rejection step to be treated classically, and one of energy registers (and acceptance-rejection qubit) to be dropped. It also has the the added benefit that generalizing this approach to Wang-Landau then allows the DOS and histogram to be treated classically as well. Finally, as the approach outlined above uses phase estimation to estimate eigenstate energies in the same way as in Ref.~\cite{temme_quantum_2011}, it inherets its ability to avoid the sign problem. These considerations lead to the quantum Wang-Landau sampling algorithm summarized in Algorithm \ref{alg:Quantum_Wang-Landau_Algorithm}.

\begin{algorithm}
\caption{Quantum Wang-Landau Algorithm}\label{alg:Quantum_Wang-Landau_Algorithm}
\KwData{Hamiltonian,\\
Number of energy bins $\ell$,\\
Initial guess for the logarithm of the relative DOS $\ln(g(E))$,\\
Initial histogram $h(E_i) = 0  \  \forall$ i,\\
Modification factor $f > 1$,\\
Reduction factor $0 < \gamma < 1$,\\
Histogram flatness criterion,\\
Total number of rounds (max number of rounds)\\}
Set round number = 0\\
Initialize the quantum computer to the state\\
$\ket{\psi}\ket{\Bar{\psi_i}}\ket{E_i}$\\
Record $E_i$, wipe all registers.\\
\While{\normalfont{round number < max number of rounds}}{
\For{ \normalfont{A predetermined number of steps}}{
Prepare the state\\
$ \frac{1}{\sqrt{N}}\sum_{x=0}^{N-1}\ket{x}\ket{x}=\frac{1}{\sqrt{N}}\sum_{i=0}^{N-1}\ket{\psi_i}\ket{\Bar{\psi_i}}$\\
Perform phase estimation and measure $E_j$\\
$\ket{\psi_i}\ket{\Bar{\psi_i}}\ket{E_j}$\\
Draw a random number\\
\uIf{\normalfont{random number} < min(1,$e^{\ln(g(E_i))-\ln(g(E_j))}$)}{
Set $\ln(g(E_j)) = \ln(g(E_j)) + \ln(f)$\\
Set $h(E_j) = h(E_j) + 1$\\
Set $E_i = E_j$\\
Wipe all registers\\
}
\Else{
Set $\ln(g(E_i)) = \ln(g(E_i)) + \ln(f)$\\
Set $h(E_i) = h(E_i) + 1$\\
Set $E_i = E_i$\\
wipe all registers\\
}}
Perform histogram check (Algorithm \ref{alg:Check_Histogram_Subalgorithm})\\
}
\end{algorithm}

\begin{figure}
\centering
\includegraphics[width=0.45\textwidth]{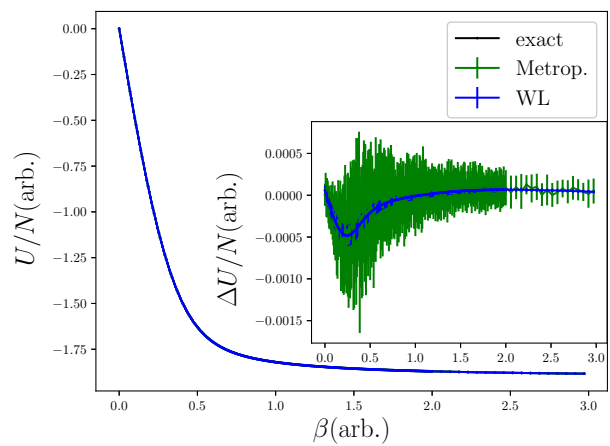}
\caption{Energy predicted by Wang-Landau, Metropolis, and exact diagonalization vs.~inverse temperature $\beta$. Errors are shown in the inset.}
\end{figure}

\begin{figure}
\centering
\includegraphics[width=0.45\textwidth]{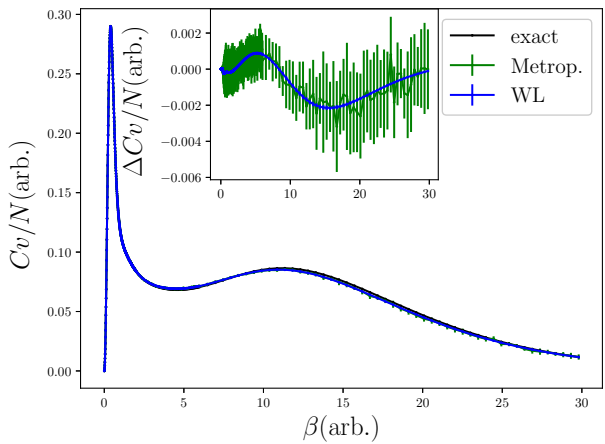}
\caption{Specific heat predicted by Wang-Landau, Metropolis, and exact diagonalization vs.~inverse temperature $\beta$.}
\end{figure}

\begin{figure}
\centering
\includegraphics[width=0.45\textwidth]{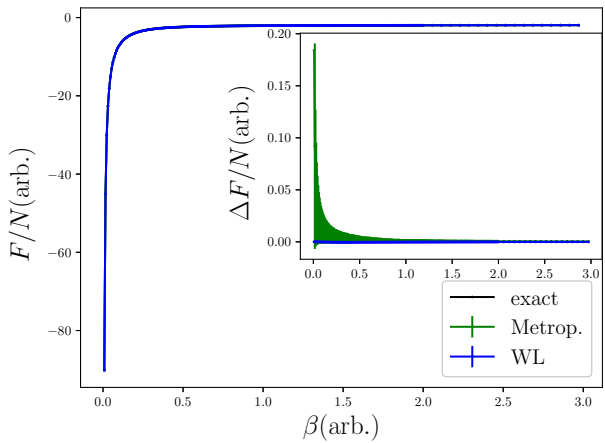}
\caption{Free energy data predicted by Wang-Landau, Metropolis, and exact diagonalization vs.~inverse temperature $\beta$.}
\end{figure}

\begin{figure}
\centering
\includegraphics[width=0.45\textwidth]{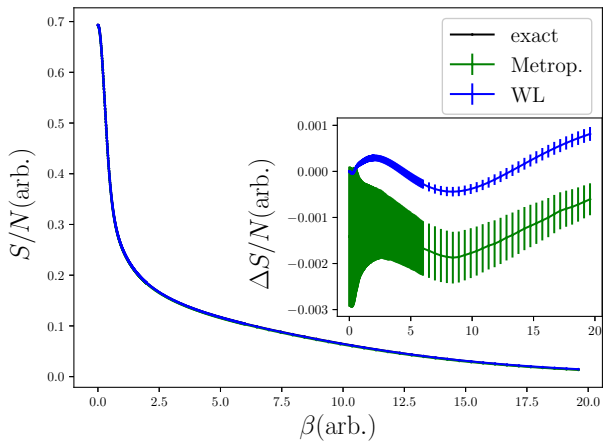}
\caption{Entropy predicted by Wang-Landau, Metropolis, and direct Hamiltonian diagonalization vs. inverse temperature $\beta$.}
\end{figure}


To assess the effectiveness of the technique, and its benefits with respect to Metropolis sampling, we perform proof of concept simulations using a 1D transverse-field Ising model with  periodic boundary conditions. In this model, a chain of quantum spins has the Hamiltonian
\begin{equation}
 H = J\sum_{i=1}^{N}\sigma^{z}_{i}\sigma^{z}_{i+1} + h\sum_{i=1}^N\sigma^{x}_{i}.
 \end{equation}
Here the summation is over the spins in the chain and because of the periodic boundary conditions the last term in the sum is $\sigma^{z}_{N}\sigma^{z}_{1}$.
Furthermore, $J$ and $h$ are constants which can be changed to evaluate model behavior for different relative interaction strengths. In this work we choose $J=2$ and $h=1$. The details for how the Monte Carlo sampling is carried out are given in the Supplemental Material.

The internal energy, heat capacity, free energy, and Gibbs entropy for 9 system qubits (spins) and 11 phase estimation qubits obtained from simulated quantum  Wang-Landau and Metropolis sampling are given in Figures 1-4. The differences between the values of each thermodynamic quantity predicted by both Monte Carlo methods and the exact value determined by Hamiltonian diagonalization are given in the insets. The error bars in each plot represent one standard deviation. The time evolution interval for phase estimation was chosen in accordance with the procedure given in the Supplemental Material. 

The Wang-Landau data are obtained from the average of twenty runs, and quantities are calculated directly from the relative DOS. The Metropolis energy and heat capacity for each temperature is calculated from the average of twenty runs with a constant number of Monte Carlo steps between runs and temperatures. The total number of Monte Carlo steps per run per temperature was chosen so that the combined number of Monte Carlo steps of all Metropolis runs would be equal to or slightly greater than that of the Wang-Landau runs after adjusting for an initial relaxation period. The first 5,000 steps of each Metropolis run were cut from data analysis to allow the system to relax to equilibrium.
 In the case of the data shown in Figures 1-4, 12,080,000 Metropolis steps were taken for each run, and all energies after 5000 steps were used to calculate thermodynamic quantities.

The energy and heat capacity given by Metropolis are calculated from taking the averages of the moments of energy. The entropy 
\begin{equation}
 S=\int_0^{T_f}\frac{C_v}{T}\,dT=\int_{\beta_f}^{\infty}\frac{C_v}{\beta}\,d\beta
 \label{entropy from heat capacity}
 \end{equation}
is calculated by evaluating the second expression in Eq.~(\ref{entropy from heat capacity}). To evaluate that integral it is necessary to truncate the upper limit; this is done by assuming that the heat capacity drops to zero one inverse temperature interval after the $\beta$ cutoff. The free energy $F=U-TS$ is calculated for each temperature from the internal energy and entropy obtained  above.

We find that there is a very small systematic error between the data given by our simulations and the exact answer given by direct diagonalization of the Hamiltonian across all observed quantities. However, excluding systematic error due to integration of Metropolis data, the observed systematic error can primarily be attributed to error in the phase estimation algorithm and can be seen to decrease with an increasing number of phase estimation qubits. Integration error in the data has the effect of systematically displacing the Metropolis entropy curve compared to that given by Wang-Landau and results in the erroneous sharp uptick in the Metropolis free energy curve at low inverse temperature compared to Wang-Landau and that given by Hamiltonian diagonalization. An analysis of the individual Wang-Landau sampling and phase estimation errors is given in the Supplemental Material.

In conclusion, we have introduced and simulated an algorithm for performing Wang-Landau sampling with a fault-tolerant quantum computer. This algorithm represents an improvement over Metropolis sampling, avoids the sign problem, and doesn't become trapped in complex free energy landscapes or at low temperature. For each of the thermodynamic quantities studied, Wang-Landau sampling is found to significantly outperform Metropolis sampling in accuracy when using the same total number of Monte Carlo steps. The improvement is especially prominent in the entropy and free energy since these cannot be directly obtained from Metropolis data and must be evaluated indirectly, increasing  sensitivity to noise. This algorithm brings the tools for simulating quantum systems to the modern era of post-Metropolis Monte Carlo sampling, enabling the simulation of complex and strongly correlated quantum matter with fault-tolerant quantum computers.

\bibliographystyle{apsrev4-2}
\bibliography{Exported_Items.bib}



\end{document}